\documentstyle[ltwol]{article}
\input BoxedEPS.tex
\SetOzTeXEPSFSpecial
\HideDisplacementBoxes 

\def\Journal#1#2#3#4{{#1} {\bf #2}, #3 (#4)}

\def\PLB{{\em Phys. Lett.}  B}
\def\PRL{\em Phys. Rev. Lett.}
\def\PRC{{\em Phys. Rev.} C}
\def\PRD{{\em Phys. Rev.} D}

\def\ra{\rightarrow}

\def\be{\begin{equation}}
\def\ee{\end{equation}}
\def\bea{\begin{eqnarray}}
\def\eea{\end{eqnarray}}
\def\roughly#1{\mathrel{\raise.3ex\hbox{$#1$\kern-.75em\lower1ex\hbox{$\sim$}}}}
\def\lsim{\roughly<}
\def\gsim{\roughly>}
\newcommand{\nsun}{\mbox{$\nu_\odot$ }}
\newcommand{\phib}{\phi(\nsun \mbox{{\rm from} }^8 {\rm B})}
\newcommand{\optbar}[1]{\shortstack{{\tiny (\rule[.4ex]{1em}{.1mm})} 
  \\ [-.7ex] $#1$}}
\newcommand{\Eq}[1]{Eq.~(\ref{eq#1})}

\begin{document}

\title{NEUTRINO MASS: WHERE DO WE STAND, AND WHERE ARE WE GOING?\footnotemark[1]}

\author{BORIS KAYSER}

\address{National Science Foundation, 4201 Wilson Blvd., Arlington, VA 22230,
USA\\E-mail: bkayser@nsf.gov}

\twocolumn[\maketitle\abstracts{We discuss the three major hints of neutrino mass: the strong evidence of mass from the behavior of atmospheric neutrinos, the very promising further hint from the behavior of solar neutrinos, and the so-far unconfirmed hint from the LSND experiment. Then, we describe two illustrative neutrino-mass scenarios suggested by the observed hints. We identify the needed future tests of both the present hints and the neutrino-mass scenarios.}]

\footnotetext[1]{To appear in the Proceedings of the 29th International Conference on High Energy Physics, Vancouver, July 1998.}  

\section{Hints of Neutrino Mass} 
\subsection{Atmospheric Neutrinos}\label{subsec:am} 
The behavior of atmospheric neutrinos, produced in the earth's
atmosphere by cosmic rays, now provides rather strong evidence that neutrinos
have mass. The atmospheric neutrinos are studied by underground detectors. A
signal that these neutrinos have mass is sought by looking for indications that
they oscillate from one flavor to another---something they can do only if they
have mass. The Super-Kamiokande underground experiment (Super~K) has found
evidence of this flavor oscillation which is particularly convincing because it
is almost independent of theoretical input.\cite{ref1} This persuasive evidence
comes from the dependence of the detected atmospheric neutrino flux on the
direction from which the neutrinos are coming.

Suppose that neutrinos do {\em not} oscillate, and consider the flux of neutrinos
of a given flavor coming from polar angle $\theta$ with respect to the
zenith---the point directly above the underground detector. Consider only
neutrinos with energies above a few GeV, so that geomagnetic effects can be
neglected. Then, given the fact that the flux of cosmic rays which produces the
neutrinos in our atmosphere is isotropic,\cite{ref2} it follows from a simple
geometric argument that, at the detector, the neutrino fluxes from $\theta$ and
$\pi - \theta$ will be equal\,\cite{ref3} to within a few percent.\cite{ref4} That
is, the neutrino flux will be up-down symmetric. 

By contrast, if neutrinos {\em do} oscillate, then in general the neutrino flux
of a given flavor will {\em not} be up-down symmetric. In the usual approximation
that only two neutrino flavors mix appreciably, the probablility that a neutrino
$\nu_f$ of flavor $f$ (e, $\mu$, or $\tau$) will turn into one of a different
flavor $f^\prime$ while traveling a distance L is given by
\begin{eqnarray}
\lefteqn{P(\nu_f \ra \nu_{f^\prime\neq f}) = } \nonumber \\
 & & \sin^2 2\alpha \sin^2 \left[ 1.27 \,\delta M^2 
({\rm eV}^2) \frac{L{\rm (km)}}{E_\nu {\rm (GeV)}} \right] .
\label{eq1}
\end{eqnarray}
Here, $E_\nu$ is the neutrino energy, $\alpha$ is a mixing angle, and $\delta M^2$ is the difference between the squared masses of the two neutrino mass eigenstates that make up $\nu_f$ and $\nu_{f^\prime}$. Now, since the Super~K detector is not far below the earth's surface, downward-going neutrinos coming from $\theta \approx 0$, and thus produced in the atmosphere directly above the detector, travel only $\sim$15 km to reach it.
In contrast, upward-going neutrinos coming from $\theta \approx\pi$, and thus produced in the atmosphere on the opposite side of the earth from the detector, travel the entire diameter of the earth, $\sim$13,000 km, to reach the detector. Let us suppose that $\delta M^2 \sim \mbox{(A Few)} \times10^{-3}$ eV$^2$, and $E_\nu \sim$ A Few GeV. Then, for the neutrinos from $\theta \approx 0$, $\delta M^2(eV^2) \left[ \frac{L{\rm (km)}}{E_\nu {\rm (GeV)}} \right] \ll1$, while for the ones from $\theta \approx \pi$, $\delta M^2({\rm eV}^2) \left[ \frac{L{\rm (km)}}{E_\nu {\rm (GeV)}} \right] >1$.
Hence, from Eq.~(\ref{eq1}), upward-going neutrinos $\nu_f$ from $\theta \approx \pi$ can oscillate away into neutrinos of a new flavor before reaching the detector, while downward-going ones from $\theta \approx 0$ cannot. This distinction will result in a difference between the detected upward- and downward-going $\nu_f$ fluxes.

Now, what do the data show? Let us consider the Super~K multi-GeV events, for which the visible energy deposition in the detector exceeds 1.33 GeV, and geomagnetic effects can be neglected. For these events, the $\nu_e$ flux is consistent with being up-down symmetric, but the $\nu_\mu$ flux has a large up-down asymmetry. To be more quantitative, let us call the flux corresponding to $-1.0 < \cos\theta < -0.2$ ``Up'', and the one corresponding to $+0.2 < \cos\theta < 1.0$ ``Down''. Then,\cite{ref4} for the $\nu_e$ events,
\be
\frac{{\rm Up}}{{\rm Down}} = 0.93^{+0.13}_{-0.12} ,  \label{eq2}
\ee
but for the $\nu_\mu$ events,
\be
\frac{{\rm Up}}{{\rm Down}} = 0.54^{+0.06}_{-0.05} .  \label{eq3}
\ee
We see that Up/Down for the $\nu_\mu$ events is very inconsistent with unity, the value it must have if there is no neutrino oscillation. Instead, Up/Down for these events suggests that the muon neutrinos which travel a significant fraction of the earth's diameter to reach the detector (the ``Up'' flux) are ocillating away into another flavor, while the ones which travel only a short distance to reach the detector (the ``Down'' flux) are not. 
Since for this data sample $\langle E_\nu\rangle =$ A Few GeV, we see from our previous discussion that a $\delta M^2 = {\cal O}(10^{-3} \: {\rm eV}^2)$ would produce this behavior. Since Up/Down for the $\nu_e$ events is consistent with unity, it appears that (at least most of) the muon neutrinos which oscillate away do not become electron neutrinos, but tau or sterile neutrinos, to which the detector is insensitive.\cite{ref5}

\begin{figure}[htb]
\center
	\vSlide{1.1in}
	\hSlide{-0.3in}
	\BoxedEPSF{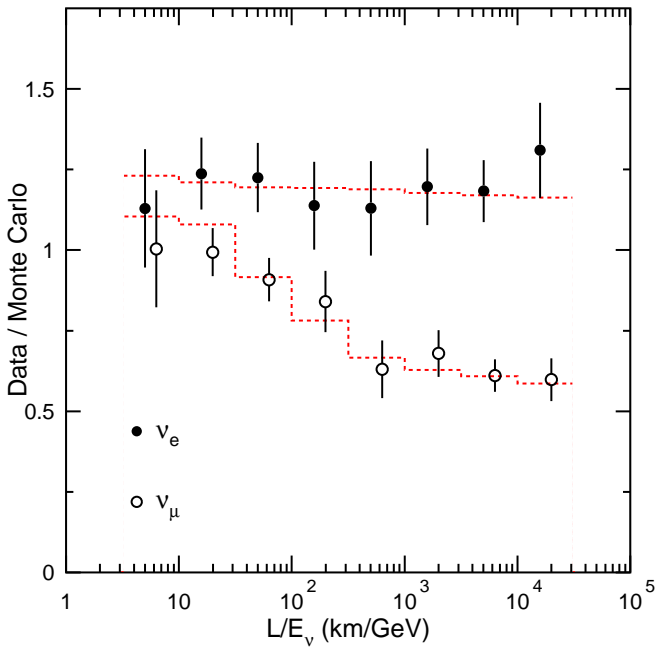  scaled 500}
\caption{Observed $\nu_e$ and $\nu_\mu$ event rates vs. $L/E_\nu$.}
\label{fig1}
\end{figure}

The Super~K detector cannot determine the L or the $E_\nu$ of an event accurately. However, estimating these quantities for each event as best as possible, the Super~K experiment finds the L/$E_\nu$ distributions shown in Fig.~\ref{fig1}.\cite{ref6} In this figure, the points show the ratios of observed event rates to Monte Carlo expectations in the absence of oscillations. 
The dashed lines show the expected shapes of these normalized event rates in the presence of $\nu_\mu \ra\nu_\tau$ oscillation with $\delta M^2 = 2.2 \times 10^{-3} \:{\rm eV}^2$ and $\sin^2 2\alpha = 1$. Note that this oscillation hypothesis fits the data very nicely.

Together, the atmospheric neutrino data from Super~K,\cite{ref1}, Soudan 2,\cite{ref7} and MACRO\,\cite{ref8} imply that the parameters of the $\nu_\mu \ra \nu_\tau$ or $\nu_\mu \ra \nu_{\rm Sterile}$ oscillation of the atmospheric neutrinos are 
\be
6 \times 10^{-4} \:{\rm eV}^2 \lsim \delta M^2 \lsim 10^{-2} \:{\rm eV}^2 \label{eq4}
\ee  and
\be  \sin^2 2\alpha \approx 1  . \label{eq5}  \ee

To confirm that the atmospheric neutrinos are oscillating, one would like to observe this same oscillation in an accelerator-generated $\nu_\mu$ beam. To do that, one will have to allow the beam to travel a distance L which is not negligible compared to the earth's diameter, since $\delta M^2 \sim 10^{-3} \:{\rm eV}^2$, and the beam energy will not be less than $\sim$1 GeV. 
During the coming years, several such ``long-baseline'' accelerator neutrino oscillation searches will be performed, using beams generated at KEK, Fermilab, and CERN. It appears that this program of experiments will be able to look for oscillation with $\delta M^2$ down to $1 \times 10^{-3} \:{\rm eV}^2$, but not lower.\cite{ref9,ref10,ref11} It is unlikely that $\delta M^2$ lies below this value.\cite{ref7,ref4,ref8}
However, it is not impossible. If $\delta M^2$ is indeed less than $1 \times 10^{-3} \:{\rm eV}^2$, then long-baseline experiments beyond those currently planned will probably be necessary. 
One intriguing possibility for an ultra-sensitive long-baseline oscillation search is to use a very intense neutrino beam which can be produced by the decay of the muons circulating in a future $\mu^+\mu^-$ collider.\cite{ref12} Indeed, if one attaches to a $\mu^+\mu^-$ collider a small muon storage ring with a straight section pointed at a suitably distant neutrino detector, one can probably extend the $\delta M^2$ reach of a $\nu_\mu \ra\nu_\tau$ search to ${\cal O}(10^{-4}\:{\rm eV}^2)$ at maximal mixing.\cite{ref13}

\vspace*{-1.8pt}   

\subsection{Solar Neutrinos}\label{subsec:sn}

Adding to the persuasive evidence of neutrino mass from the atmospheric neutrinos, the behavior of the solar neutrinos provides a strong further hint that neutrinos have nonzero masses. The observed solar neutrino ($\nu_\odot$) fluxes are all below Standard Solar Model (SSM) predictions.\cite{ref14} 
It has proved very difficult to explain these flux deficits by modifying the solar and/or nuclear physics assumed by the SSM, without invoking neutrino mass.\cite{ref15,ref16} We would like to give just one example of this difficulty: Owing to the relatively high \nsun energy threshold of the Super~K detector, the only solar neutrinos seen by this detector are from $^8$B decay. The Super~K experiment finds that\cite{ref17}
\be
\frac{\left. \phib \right|_{\rm Observed}}
     {\left. \phib \right|_{\rm SSM}} = 0.47 \pm 0.08  .
\label{eq6}
\ee
Now, the \nsun flux produced by $^8$B decay depends very sensitively on T$_{\rm core}$, the temperature in the core of the sun: $\phib \propto$ T$_{\rm core}^{24}$. Thus, it is tempting to suppose that the low observed $^8$B flux may be due simply to a T$_{\rm core}$ slightly below that in the SSM. 
However, helioseismology now tells us that T$_{\rm core} \left.\geq 0.998 \:
{\rm T}_{\rm core} \right|_{\rm SSM}$.\cite{ref18} Using $\phib \propto$ T$_{\rm core}^{24}$, it follows that  $\phib \geq 0.95 \left. \phib \right|_{\rm SSM}$. That is, by simply lowering T$_{\rm core}$ from its SSM value by the tiny amount allowed by helioseismology, one cannot come close to explaining the $\sim 50\%$ shortfall in the observed $^8$B flux.

By contrast, if we do invoke neutrino mass, then the observed \nsun fluxes can be explained successfully and elegantly. If neutrinos have mass, then a $\nu_e$---the type of neutrino generated in the center of the sun, and to which present solar neutrino detectors are sensitive---can convert into another type of neutrino, $\nu_x$, to which these detectors are completely, or at least largely, insensitive. The $\nu_x$ may be a  $\nu_\mu$,  $\nu_\tau$, or sterile neutrino $\nu_S$. 
Perhaps the most appealing $\nu_e \ra\nu_x$ conversion mechanism is the Mikheyev, Smirnov, Wolfenstein (MSW) effect. In this effect, the $\nu_e \ra\nu_x$ conversion results from the crossing, somewhere in the sun, of the total energy levels of  $\nu_e$ and  $\nu_x$.
Since the  $\nu_e$ has a charged-current interaction with solar electrons, its total energy in the sun includes not only its free-particle energy, $\sqrt{p_\nu^2 + M_{\nu_e}^2}$, but also a $\nu_e - e$ interaction energy, $\sqrt{2}\, G_F N_e$. Here, $p_{\nu}$ is the neutrino momentum, $M_{\nu_e}$ is the mass of $\nu_e$, $G_F$ is the Fermi constant, and $N_e$ is the solar electron density. 
In contrast to the $\nu_e$, the $\nu_{x(=\mu, \tau, {\rm or}\: S)}$ does not have a charged-current interaction with electrons, so its total energy in the sun, for the given $p_\nu$, is simply $\sqrt{p_\nu^2 + M_{\nu_x}^2}$. The level crossing condition is then
\be
  \sqrt{p_\nu^2 + M_{\nu_x}^2} = \sqrt{p_\nu^2 + M_{\nu_e}^2} + \sqrt{2}\, G_F N_e .
\label{eq7}
\ee
In the sun, $p_\nu \sim 1$ MeV and $N_e$ is characteristically $10^{26}$/cc, so for this condition to be satisfied, we require a mass splitting
\be
  M_{\nu_x}^2 - M_{\nu_e}^2 \equiv \delta M_{\nu_x \nu_e}^2 \sim 10^{-5}\:
  {\rm eV}^2  .
\label{eq8}
\ee
It is found that when the MSW effect is invoked, there are two regions in $(\sin^2
2\alpha_{\nu_x \nu_e}, \delta M_{\nu_x \nu_e}^2)$ space that fit the \nsun data.\cite{ref19} One has $\sin^2 2\alpha_{\nu_x \nu_e} \sim 4 \times 10^{-3}$ and the other $\sin^2 2\alpha_{\nu_x \nu_e} \sim 0.8$. For the reason just explained, both have $\delta M_{\nu_x \nu_e}^2 = {\cal O}(10^{-5}\: {\rm eV}^2)$.

The \nsun data can also be explained assuming that it is two-flavor neutrino oscillation in vacuum, taking place between the sun and the earth and described by \Eq{1}, that causes the $\nu_e \ra\nu_x$ conversion of the solar neutrinos.\cite{ref20,ref16}
To account in this manner for the event rates observed in all the \nsun experiments, one requires that (cf. \Eq{1})
\be
  \delta M^2 \frac{L}{E_\nu} \sim 1  ,  \label{eq9}
\ee
where now L is the distance from the sun to the earth, and $E_\nu \sim 1$ MeV is the energy of a typical solar neutrino. One may feel that for $\delta M^2$ to be such that \Eq{9} is satisfied when L is the distance from the sun to the earth demands a somewhat implausible coincidence. Be that as it may, the $\delta M^2$ that does the job is
\be
  \delta M^2 \sim 10^{-(9 \!-\! 11)} \,{\rm eV}^2  .   \label{eq10}
\ee
To fit the data, the mixing angle is
\be
  0.7 \lsim \sin^2 2\alpha \lsim 1.0  .                \label{eq11}
\ee

If, either through the MSW effect or oscillation in vacuum, solar electron neutrinos are being converted into muon or tau neutrinos, then obviously one would like to confirm the presence of these muon or tau neutrinos in the solar neutrino flux. An experiment which will be able to see them is being constructed in Canada. 
However, it is possible that the solar electron neutrinos are being converted, not to muon or tau neutrinos, but to sterile ones. Thus, the presence or absence of a $\nu_\mu$ or $\nu_\tau$ component in the solar flux is not a completely sure-fire test of solar neutrino flavor conversion.

Whether such conversion occurs via the MSW effect or vacuum oscillation, its probability is, in general, energy dependent. Thus, it distorts the energy spectrum of solar electron neutrinos which retain their original flavor. Clearly, one would like to observe this distortion. The Super~K experiment has already obtained some early information on the energy spectrum.\cite{ref17} It will be interesting to see what further spectral studies will show.

Additional confirmation of solar neutrino flavor conversion would come from observation of a day/night or seasonal variation in the solar $\nu_e$ flux. The ``day/night effect'' could be greatly enhanced for neutrinos passing through the earth's core.\cite{ref22}

\vspace*{-1.8pt}   

\subsection{LSND}\label{subsec:lsnd}

A third, so far unconfirmed, hint of neutrino mass comes from the Liquid Scintillator Neutrino Detector (LSND) at Los Alamos. Studying the $\bar{\nu}_\mu$ flux from $\mu^+$ decay at rest, LSND finds evidence for the appearance of $\bar{\nu}_e$, signalling the occurrence of the oscillation $\bar{\nu}_\mu \ra \bar{\nu}_e$.\cite{ref23} Studying the $\nu_\mu$ flux from the decay $\pi^+ \ra \mu^+ \nu_\mu$ of positive pions in flight, this experiment finds evidence for the appearance of $\nu_e$, signalling the occurrence of $\nu_\mu \ra \nu_e$.\cite{ref24} 
Since $L/E_\nu \sim 30$\,m/30\,MeV in LSND, we see from \Eq{1} that we must have $\delta M^2 \gsim 1$\,eV$^2$ to account for the observed oscillations. More specifically, the $\optbar{\nu}_\mu \ra \optbar{\nu}_e$ interpretation of the LSND data favors
\be
  0.2 \,{\rm eV}^2\lsim \delta M^2 \lsim 10 \,{\rm eV}^2   \label{eq12}
\ee   and  \be
  0.002 \lsim \sin^2 2\alpha \lsim 0.03.  \label{eq13}
\ee

A number of other experiments have placed upper bounds on $\optbar{\nu}_\mu \ra \optbar{\nu}_e$ oscillation.\cite{ref25} Together, the KARMEN~2 and Bugey bounds allow only a narrow sliver of the $(\delta M^2, \sin^2 2\alpha)$ region favored by LSND. However, KARMEN~2 anticipated seeing three background events, and in fact saw no events---background or signal. 
Had KARMEN~2 seen the expected three background events, this experiment, plus all the others with bounds, would have allowed a good deal of the LSND-favored region. Thus, the situation with respect to the LSND oscillation signal appears to be somewhat unsettled. It is important to confirm or disprove the $\optbar{\nu}_\mu \ra \optbar{\nu}_e$ interpretation of the LSND data through future studies. These studies will be carried out both by KARMEN and by a new experiment at Fermilab.

\section{Hot Dark Matter}

Evidence is accumulating that
\be  \Omega_M = 0.1 - 0.4,  \label{eq14}  \ee
rather than unity.\cite{ref26,ref27} Here, $\Omega_M$ is the total mass density of the universe, as a fraction of the critical density required to stop indefinite expansion. Now, much of the $\Omega_M$ of \Eq{14} is dark, and some of this dark matter {\em might} be Hot Dark Matter (HDM); that is, dark matter consisting of particles whose masses are small compared to 1 keV. 
Neutrinos are natural candidates for such particles. However, some of the dark matter must be Cold Dark Matter (CDM), consisting of particles much heavier than neutrinos, because these heavier particles were needed to seed the early-universe formation of the observed galactic structure. 
If $\Omega_M \lsim 0.4$, and the mass density of CDM is, say, at least 0.2, then the mass density of HDM, $\Omega_{HDM}$, obeys the constraint
\be  \Omega_{HDM} \lsim 0.2  .  \label{eq15}  \ee

The relation between the masses $M_{\nu_m}$ of the neutrino mass eigenstates $\nu_m$ and the mass density $\Omega_\nu$ in neutrinos is
\be
  M_{\nu_1} + M_{\nu_2} + M_{\nu_3} = \Omega_\nu (33 \, {\rm eV}) .  \label{eq16}
\ee
Here, we are assuming that there are only three mass eigenstates, which make up $\nu_e$, $\nu_\mu$, and $\nu_\tau$. Now, if neutrinos are light compared to 1 keV, as certainly appears to be the case, then $\Omega_\nu \leq \Omega_{HDM}$. Then, Eqs.~(\ref{eq15}) and (\ref{eq16}) imply that
\be
  M_{\nu_1} + M_{\nu_2} + M_{\nu_3} \lsim 6 \, {\rm eV} .  \label{eq17}
\ee
This condition is not a hint of neutrino mass, but a {\em constraint} on neutrino mass.

\section{Where Does the Evidence Point?}
We have discussed the evidence for neutrino mass from the behavior of atmospheric neutrinos, the major further hint of mass from the behavior of solar neutrinos, and the unconfirmed additional hint from the LSND experiment. To what neutrino mass scenarios do these hints point?

Suppose that the atmospheric, solar, and LSND neutrinos all oscillate. The observed ``frequencies'' $\delta M^2$ of these three oscillations cannot all be accommodated in any simple way in terms of the masses of just the three mass eigenstate neutrinos making up $\nu_e$, $\nu_\mu$, and $\nu_\tau$. 
If there are only three mass eigenstate neutrinos, $\nu_1$, $\nu_2$, and $\nu_3$, then there are just three different splittings $\delta M^2_{\nu_i \nu_j} \equiv M^2_{\nu_i} - M^2_{\nu_j}$, and these three splittings obviously satisfy
\bea
  \sum_{\rm Splittings} \delta M^2_{\nu_i \nu_j} & = & (M^2_{\nu_3} - M^2_{\nu_2}) 
  + (M^2_{\nu_2} - M^2_{\nu_1}) \nonumber \\
  & & \hspace{.5in} + (M^2_{\nu_1} - M^2_{\nu_3}) \nonumber \\
  & = & 0  .		\label{eq18}
\eea 
But, as we have seen, the splittings required by the three hints of oscillation are approximately as summarized in Table 1.
\begin{table}[htb]
\begin{center}
\caption{Mass splittings required by the three hints of oscillation.}
\begin{tabular}{|c|c|} 
\hline 
\raisebox{0pt}[12pt][6pt]{Oscillating Neutrinos} & 
  \raisebox{0pt}[12pt][6pt]{Required $\left| \delta M^2 \right|$\,(eV$^2$)} \\
\hline
\raisebox{0pt}[12pt]{Solar} & 
  \raisebox{0pt}[12pt]{$10^{-10}$ or $10^{-5}$} \\
Atmospheric & $10^{-3}$ to $10^{-2}$ \\
LSND & $10^{-1}$ to $10^{+1}$ \\ 
\hline
\end{tabular}
\end{center}
\vspace*{3pt}
\end{table}
Since the $\delta M^2$ values required by the three hints are, respectively, of three different orders of magnitude, there is no way they can add up to zero, as demanded by \Eq{18}, no matter which sign we assign to each of them. Thus, just three neutrinos cannot explain the observed three hints of oscillation.\cite{ref28} 
To accommodate all three of these hints, we must introduce (at least) a fourth neutrino. Since we know from the width of the weak boson $Z^0$ that only three species of neutrinos have normal electroweak interactions, the extra, fourth neutrino must be a sterile neutrino $\nu_S$.

\begin{figure}[htb]
\center
	\BoxedEPSF{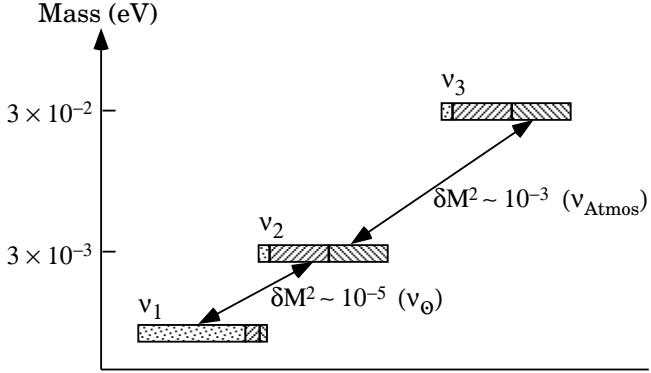  scaled 700}
\caption{A three-neutrino mass hierarchy that accounts for the oscillations of atmospheric and solar neutrinos.}
\label{fig2}
\end{figure}

If one is uneasy about introducing a light sterile neutrino, then he can proceed by provisionally setting aside the so-far unconfirmed LSND result, and attempting to explain only the atmospheric and solar neutrino results in terms of neutrino oscillation. One can then have the neutrino mass spectrum shown in Fig.~\ref{fig2}.
There, $\nu_1$, $\nu_2$, and $\nu_3$ are the mass eigenstates. The flavor content of each of these is indicated by dotting and hatching: The $\nu_e$ fraction of a mass eigenstate is dotted, the $\nu_\mu$ fraction is indicated by right-leaning hatching, and the $\nu_\tau$ fraction by left-leaning hatching.\cite{ref29} 
The masses in Fig.~\ref{fig2} form a hierarchy: $M_{\nu_3} \gg M_{\nu_2} \gg M_{\nu_1}$. The mass-squared splitting $M^2_{\nu_3} - M^2_{\nu_2} \simeq M^2_{\nu_3} \simeq 10^{-3}$\,eV$^2$ gives the atmospheric neutrino oscillation, while the splitting $M^2_{\nu_2} - M^2_{\nu_1} \simeq M^2_{\nu_2} \simeq 10^{-5}$\,eV$^2$ yields the MSW effect in the sun.

The hierarchy in Fig.~\ref{fig2} is the lightest one which gives the $\delta M^2$ values required by the atmospheric and solar oscillations. Since oscillations determine only mass splittings, and not actual masses, one, of course, can increase the masses in  Fig.~\ref{fig2} without affecting the oscillations so long as one keeps the $\delta M^2$ values fixed. 
For example, if neutrinos contribute significantly to $\Omega_M$, then perhaps $M_{\nu_1} \simeq M_{\nu_2} \simeq M_{\nu_3} \simeq (1-2)$\,eV, with the $\delta M^2$ values as in Fig.~\ref{fig2}.

\begin{figure}[htb]
\center
	\tBoxedEPSF{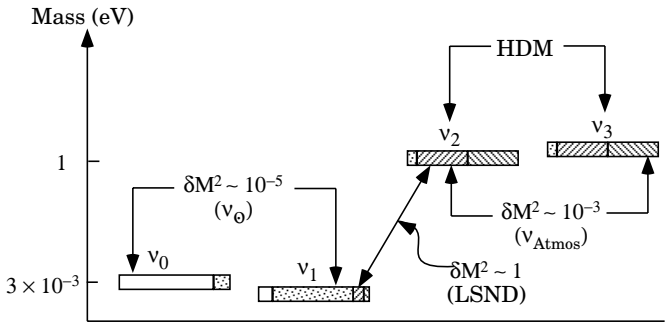  scaled 600}
\caption{A four-neutrino scenario that accounts for the oscillations of the atmospheric, solar, and LSND neutrinos. In this scenario, the mass eigenstate $\nu_0$ is largely a sterile neutrino $\nu_S$. The $\sim 1$ eV neutrinos in the scheme make a significant contribution to Hot Dark Matter.}
\label{fig3}
\end{figure}

If one accepts the inclusion of a $\nu_S$ among the light neutrinos, then all three hints of oscillation can be accommodated. One neutrino mass spectrum which does this is shown in Fig.~\ref{fig3}.\cite{ref30}
In this spectrum there are, of course, four mass eigenstates, $\nu_0 - \nu_3$. The $\nu_e$, $\nu_\mu$, and $\nu_\tau$ fractions of these mass eigenstates are indicated as they were in Fig.~\ref{fig2}, and the $\nu_S$ fraction is shown as a clear region. 
There are two $\sim 1$\,eV mass eigenstates, $\nu_2$ and $\nu_3$, with a mass splitting $\delta M^2 \sim 10^{-3}$\,eV$^2$ that yields the atmospheric neutrino oscillation. Then there are two much lighter mass eigenstates, $\nu_0$ and $\nu_1$, with a splitting $\delta M^2 \sim 10^{-5}$\,eV$^2$ that gives the MSW effect in the sun. 
However, from Fig.~\ref{fig3} we see that in this scheme the MSW effect converts a $\nu_e$ into a {\em sterile} neutrino, rather than a muon or tau neutrino. Thus, there is no $\nu_\mu$ or $\nu_\tau$ component in the solar neutrino flux. Finally, the splitting $\delta M^2 \sim 1$\,eV$^2$ between the heavy pair $\nu_2$ and $\nu_3$, and the light pair $\nu_0$ and $\nu_1$, produces the LSND oscillation. 
The heavy pair in this scenario has the smallest average mass, $\sim 1$\,eV, which can still yield the $\delta M^2 \sim 1$\,eV$^2$ called for by LSND. Interestingly, this $\sim 1$\,eV mass has the consequence that the members of this pair make a significant hot dark matter contribution to the mass density of the universe, but without violating the cosmological upper bound of \Eq{17} on neutrino mass.

In theories of neutrino mass, sterile neutrinos occur commonly and naturally. However, these sterile neutrinos tend to be heavy, with masses in the multi-GeV range or higher. {\em Light} sterile neutrinos, with masses of 1 eV or less, are not generally anticipated. Confirmation of the existence of a light sterile neutrino such as the one required by the scenario of Fig.~\ref{fig3} would be as groundbreaking a finding as the original discovery of neutrino mass. 
Thus, it is very important to carry out the future experiments which will test the oscillation interpretations of the atmospheric, solar, and LSND results, and to thoroughly examine the question of whether or not these three oscillations, if all confirmed, together require the existence of a fourth, sterile neutrino.

The two neutrino-mass scenarios we have discussed are just examples. Other scenarios are possible. The candidate scenarios can be tested through future experiments. For example, the three-neutrino scheme of Fig.~\ref{fig2} predicts a $\nu_{\mu,\tau}$ component in the solar neutrino flux, while the four-neutrino scheme of Fig.~\ref{fig3} predicts none. The four-neutrino scheme predicts the confirmation of LSND, while the three-neutrino one predicts its disproof. Both scenarios predict a distortion of the energy spectrum of electron neutrinos from the sun, and other phenomena as well.

\section{Conclusion}

Thanks to the strong atmospheric-neutrino evidence that neutrinos do have nonzero masses, this is without question an exciting time in neutrino physics. What we need now are further strong signals that neutrinos have mass, and more insight---experimental and theoretical---into the masses and mixings chosen by nature.


\section*{Acknowledgements}
The author is grateful for the hospitality of the Aspen Center for Physics, where it was possible to learn a good deal about the recent developments in neutrino physics through conversations with many people.


\section*{References}
\begin{thebibliography}{99}

\bibitem{ref1}Y. Fukuda {\it et al.} (Super-Kamiokande Collaboration), \Journal{\PRL} {81}{1562}{1998}.
\bibitem{ref2}We thank Donald Groom and Eugene Loh for conversations on this point.
\bibitem{ref3}We thank Tom Gaisser and John Learned for discussions of this point.
\bibitem{ref4}T. Kajita, talk presented at Neutrino 98, held in Takayama, Japan, June 1998.
\bibitem{ref5}This conclusion is strengthened by the CHOOZ reactor limit on $\nu_\mu \ra \nu_e$ presented in M. Apollonio {\it et al.}, \Journal{\PLB}{420}{397}{1998}.
\bibitem{ref6}We thank James Stone and Ed Kearns for permission to reproduce this figure from Ref.~1.
\bibitem{ref7}H. Gallagher, talk presented at this conference.
\bibitem{ref8}D. Michael,  talk presented at this conference.
\bibitem{ref9}K. Nishikawa, talk presented at Neutrino 98.
\bibitem{ref10}S. Wojcicki, talk presented at Neutrino 98.
\bibitem{ref11}F. Pietropaolo, talk presented at Neutrino 98.
\bibitem{ref12}For a discussion of neutrino oscillation experiments with neutrino beams from a muon collider, see the {\em Proceedings of the Workshop on Physics at the First Muon Collider and at the Front End of the Muon Collider}, edited by S. Geer and R. Raja (American Inst. of Phys., Woodbury, NY, 1998).
\bibitem{ref13}This is a guess based on the related $\nu_e \ra \nu_{\mu,\tau}$ sensitivities reported by S. Geer in \Journal{\PRD}{57}{6989}{1998} and in Ref.~12.
\bibitem{ref14}Predictions of the most recent version of the SSM may be found in J. Bahcall, S. Basu, and M. Pinsonneault, \Journal{\PLB}{433}{1}{1998}.
\bibitem{ref15}J. Bahcall and H. Bethe, \Journal{\PRL}{65}{2233}{1990} and \Journal{\PRD}{44}{2962}{1991}.
\bibitem{ref16}N. Hata and P. Langacker, \Journal{\PRD}{56}{6107}{1997}.
\bibitem{ref17}Y. Suzuki, talk presented at Neutrino 98.
\bibitem{ref18}J. Bahcall, M. Pinsonneault, S. Basu, and J. Christensen-Dalsgaard, \Journal{\PRL}{78}{171}{1997}.
\bibitem{ref19}We thank N. Hata and P. Langacker for sharing their recent fits with us.
\bibitem{ref20}P. Krastev and S. Petcov, \Journal{\PRD}{53}{1665}{1996}.
\bibitem{ref22}S. Petcov, \Journal{\PLB}{434}{321}{1998}; M. Chizhov, M. Maris and S. Petcov, SISSA preprint 53/98/EP; V. Ermilova, V. Tsarev, and V. Chechin, \Journal{{\em Short Notices of the Lebedev Inst.}}{5}{26}{1986}; E. Akhmedov, \Journal{{\em Sov. J. Nucl Phys.}}{47}{301}{1988}; P. Krastev and A. Smirnov, \Journal{\PLB}{226}{341}{1989}.
\bibitem{ref23}C. Athanassopoulos {\it et al.} (LSND Collaboration), \Journal{\PRC}{54}{2685}{1996} and \Journal{\PRL}{77}{3082}{1996}.
\bibitem{ref24}C. Athanassopoulos {\it et al.} (LSND Collaboration), \Journal{\PRL}{81}{1774}{1998}.
\bibitem{ref25}Notable among these bounds is one from the KARMEN~2 experiment, discussed by J. Kleinfeller in a talk at this conference. Other bounds come from the Bugey reactor-neutrino experiment, and the NOMAD and CCFR accelerator-neutrino experiments.
\bibitem{ref26}N. Bahcall and X. Fan, \Journal{{\em Proc. Natl. Acad. Sci. USA}}{95}{5956}{1998}.
\bibitem{ref27}We thank Gary Steigman for a useful discussion and Neta Bahcall for several illuminating lectures on the mass density of the universe.
\bibitem{ref28}This argument assumes that each of the three oscillations may be described by just a single $\delta M^2$. There have been interesting attempts (R. Thun and S. McKee, hep-ph/9806534; T. Teshima and T. Sakai, hep-ph/9805386; 
C. Cardall and G. Fuller, talk presented by C. Cardall at the 1998 Aspen Winter Conference on Particle Physics, held in Aspen, CO, January 1998) to fit all three oscillations with just three neutrinos by dropping this simple assumption. However, it appears to us that every such attempt is in some conflict with {\em some} of the oscillation data.
\bibitem{ref29}We owe this general approach to depicting neutrino-mass scenarios to A. Smirnov, in {\em Proc. 28th Int. Conf. on High Energy Physics,} edited by Z. Ajduk and A. Wroblewski (World Scientific, Singapore, 1997) p.288.
\bibitem{ref30}This is a somewhat modified version of a neutrino-mass scenario proposed in D. Caldwell and R. Mohapatra, \Journal{\PRD}{48}{3259}{1993}. See also V. Barger, S. Pakvasa, T. Weiler, and K. Whisnant, hep-ph/9806328.

\end{thebibliography}
\end{document}